\documentclass[modern]{aastex62}
\usepackage[margin=1in]{geometry}
\usepackage{verbatim}
\usepackage{sidecap}
\renewcommand{\baselinestretch}{0.95}
\NewPageAfterKeywords

\newcommand{\WF}{{\em WFIRST}}
\newcommand{\Kep}{{\em Kepler}}

\defcitealias{Suzuki16}{S16}
\defcitealias{Jung18}{J18}

\begin{document}

\begin{flushleft}
{\large Astro2020 Science White Paper} 
\linebreak

{\bf\large The Scientific Context of \WF\, Microlensing in the 2020s}
\linebreak
\normalsize

\noindent \textbf{Thematic Areas:} \hspace*{60pt} $\checkmark$ Planetary Systems \hspace*{10pt} $\checkmark$ Star and Planet Formation \hspace*{20pt}\linebreak
$\square$ Formation and Evolution of Compact Objects \hspace*{31pt} $\square$ Cosmology and Fundamental Physics \linebreak
  $\square$  Stars and Stellar Evolution \hspace*{1pt} $\square$ Resolved Stellar Populations and their Environments \hspace*{40pt} \linebreak
  $\square$    Galaxy Evolution   \hspace*{45pt} $\square$             Multi-Messenger Astronomy and Astrophysics \hspace*{65pt} \linebreak
  
\textbf{Principal Author:}

Name: Jennifer C. Yee 
 \linebreak  				
Institution: Center for Astrophysics | Harvard \& Smithsonian, 60 Garden St. MS-15, Cambridge, MA 02138 
 \linebreak
Email: jyee@cfa.harvard.edu
 \linebreak
Phone: 617-495-7594 
 \linebreak
 
\textbf{Co-authors:} (names and institutions)
  \linebreak
Rachel Akeson (Caltech/IPAC),
Jay Anderson (Space Telescope Science Institute),
Etienne Bachelet (Las Cumbres Observatory),
Charles Beichman (NASA Exoplanet Science Institute),
Andrea Bellini (Space Telescope Science Institute),
David P. Bennett (NASA Goddard \& UMD),
Aparna Bhattacharya (NASA Goddard \& University of Maryland College Park ),
Valerio Bozza (Università di Salerno),
Geoffrey Bryden (Jet Propulsion Laboratory),
Sebastiano Calchi Novati (Caltech/IPAC),
B. Scott Gaudi (The Ohio State University),
Andrew Gould (The Ohio State University),
Calen B. Henderson (Caltech/IPAC),
Savannah R. Jacklin (Vanderbilt University),
Samson A. Johnson (The Ohio State University),
Naoki Koshimoto (The University of Tokyo \& National Astronomical Observatory of Japan),
Shude Mao (Tsinghua University),
David M. Nataf (The Johns Hopkins University),
Matthew Penny (The Ohio State University),
Radoslaw Poleski (The Ohio State University),
Cl\'ement Ranc (NASA/Goddard Space Flight Center),
Kailash Sahu (Space Telescope Science Institute),
Yossi Shvartzvald (Caltech/IPAC),
Keivan G. Stassun (Vanderbilt University),
Rachel Street (Las Cumbres Observatory)

\end{flushleft}

\begin{abstract}
As discussed in {\em Exoplanet Science Strategy} \citep{NAS_ESS18}, \WF\, \citep{WFIRST2020s} is uniquely capable of finding planets with masses as small as Mars at separations comparable to Jupiter, i.e., beyond the current ice lines of main sequence stars. In semimajor axis, these planets fall between the close-in planets found by \Kep\, \citep{Coughlin16} and the wide separation gas giants seen by direct imaging \citep[e.g. ][]{Lagrange09} and ice giants inferred from ALMA observations \citep{Zhang18}. Furthermore, the smallest planets \WF\, can detect are smaller than the planets probed by radial velocity \citep{Mayor11, Bonfils13} and {\em Gaia} \citep{Perryman14} at comparable separations.
Interpreting planet populations to infer the underlying formation and evolutionary processes requires combining results from multiple detection methods to measure the full variation of planets as a function of planet size, orbital separation, and host star mass. Microlensing is the only way to find planets from $0.5$ to $5 M_{\oplus}$ at separations of $1$ to $5$ au.

Fundamentally, the case for a microlensing survey from space has not changed in the past 20 years: going to space allows wide-field diffraction-limited observations that can resolve main-sequence stars in the bulge, which in turn allows the detection and characterization of the smallest microlensing signals including those from planets with masses at least as small as Mars {\citep{BennettRhie02}}. What has changed is that ground-based microlensing is reaching its limits, which underscores the scientific necessity for a space-based microlensing survey to measure the population of the smallest planets. Ground-based microlensing has found a break in the mass-ratio distribution at about a Neptune mass-ratio \citep{Suzuki16, Jung18}, implying that Neptunes are the most common microlensing planet and that planets smaller than this are rare.  However, ground-based microlensing reaches its detection limits at mass ratios only slightly below the observed break. The \WF\, microlensing survey will measure the shape of the mass-ratio function below the break by finding numerous smaller planets: $\sim 500$ Neptunes, a comparable number of large gas giants, and $\sim 200$ Earths (if they are as common as Neptunes), and it can detect planets as small as $0.1 M_{\oplus}$ \citep{Penny18}. In addition, because it will also measure host star masses and distances, \WF\, will also track the behavior of the planet distribution as a function of separation and host star mass.
\end{abstract}

\pagebreak
\section{Introduction}

Because microlensing is the only planet-finding technique that probes planets as small as Mars just beyond the current ice lines of their hosts, it is essential for a complete sample of exoplanet populations \citep{NAS_ESS18}. Thus, there is a broad case for \WF\, microlensing, which will detect $\gtrsim$ a thousand planets. A sample of this size will allow direct tests of planet formation theory, novel discoveries, and provide context for understanding the planets of our solar system \citep[see e.g.,][]{Bennett10_MPF, Spergel15, NAS_ESS18}.

Rather than providing a broad perspective on the \WF\, microlensing science case, the purpose of this white paper is to explore in detail a few specific, well-defined, science questions motivated by recent research that can be answered by \WF. 

\section{Questions that Space-Based Microlensing Can Answer
\label{sec:questions}}

\begin{figure}
\includegraphics[height=0.3\textheight]{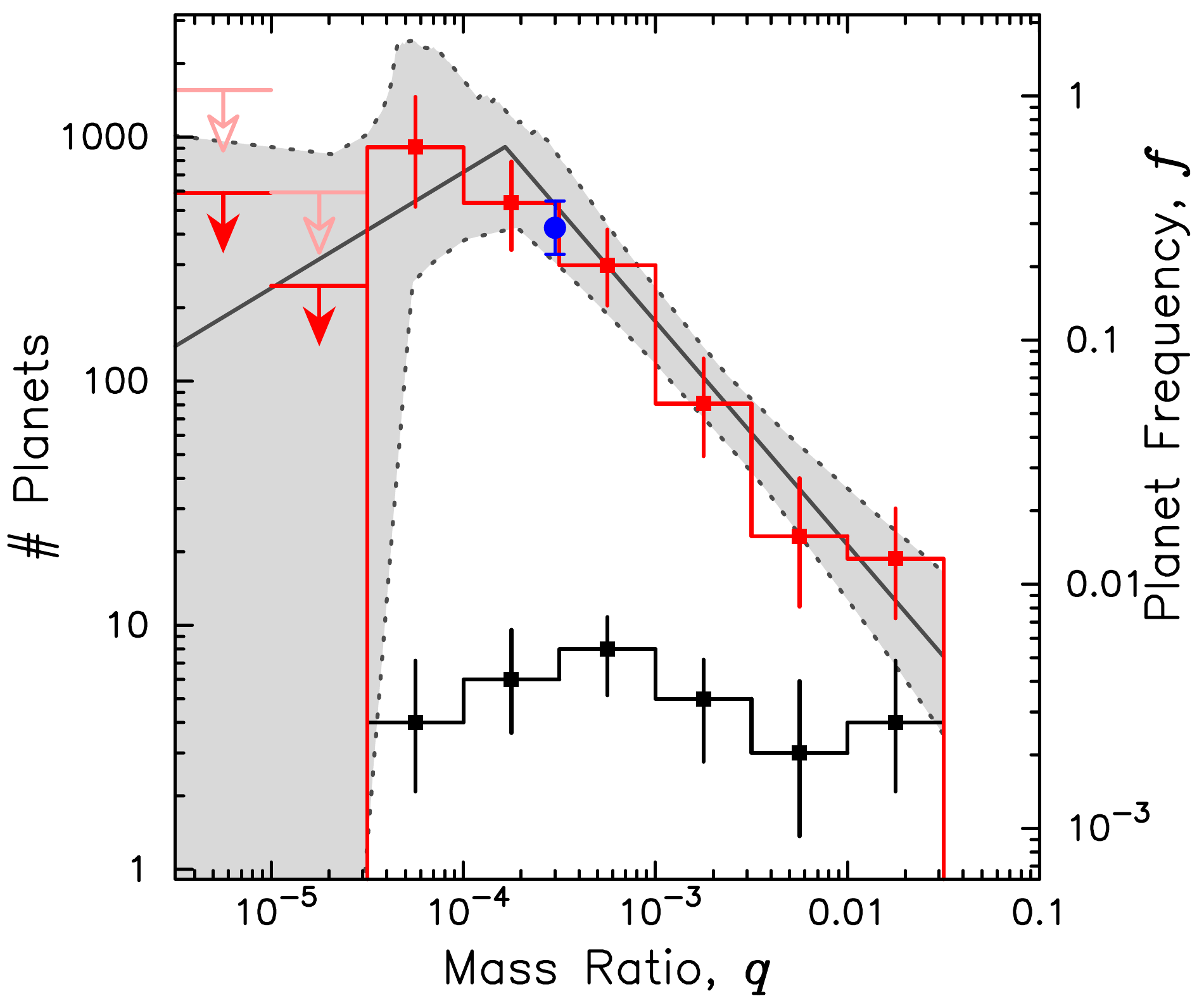}
\includegraphics[height=0.3\textheight]{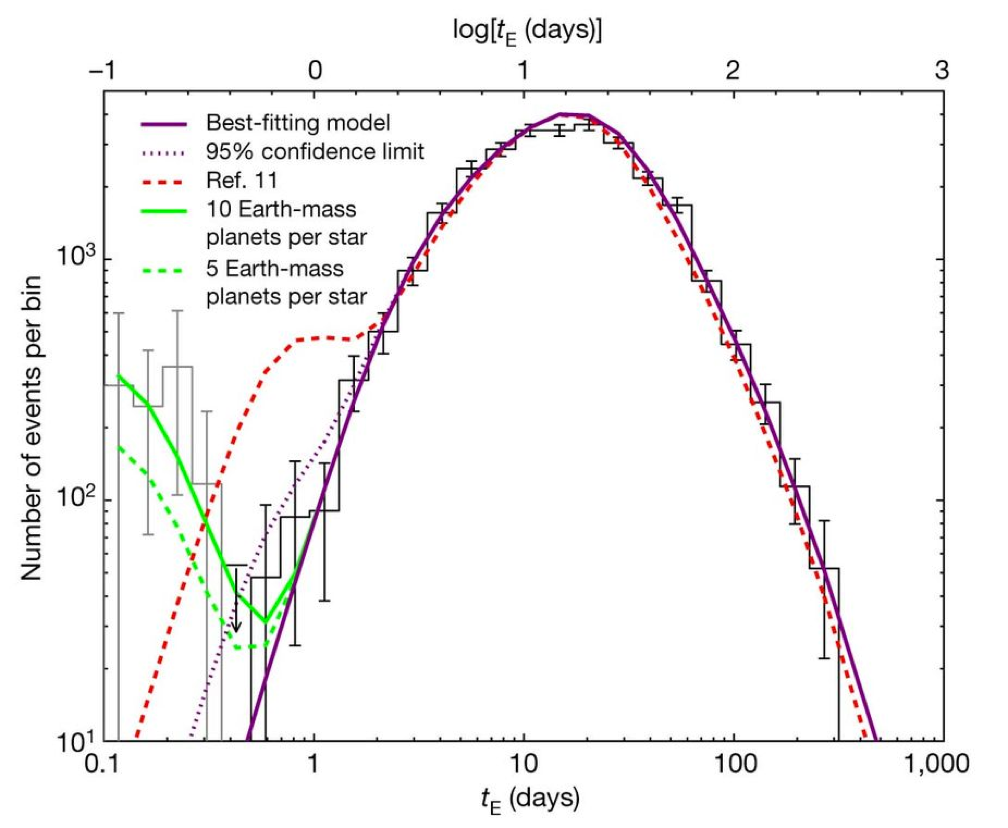}
\caption{ 
{\it Left}: the planet mass-ratio distribution measured by \citetalias{Suzuki16} shows a peak that \citetalias{Jung18} constrain to be at $q_{\rm br} \sim 5\times10^{-5}$, but ground-based microlensing is only marginally sensitive to planets below $q_{\rm br}$ \citepalias[figure from ][]{Suzuki16}.
{\it Right}: the timescale distribution of single microlenses measured by OGLE \citep{Mroz17FFPs} ({\it solid, purple} line) is consistent with no free-floating Jupiters which rules out the \citet{Sumi11} ({\it dashed, red} line) estimate of 2 free-floating Jupiters per star \citep[figure from ][]{Mroz17FFPs}. However, there is tentative evidence for a population of {smaller} free-floating planets ({\it gray} and {\it green} distributions), which, {if real}, can be characterized by \WF.
\label{fig:gbulens}}
\end{figure}

\subsection{What is the Shape of the Mass-Ratio Distribution of Planets?}

Ground-based microlensing shows the mass-ratio ($q = m_{\rm p} / M_{\rm star}$) distribution of exoplanets at a few au can be described by a broken power law that peaks at a Neptune mass ratio, i.e.,  $q_{\rm br} \sim q_{\rm Nep} \sim 5 \times 10^{-5}$ \citep[Figure \ref{fig:gbulens};][hereafter S16 and J18, respectively]{Suzuki16, Jung18}. For planets with $q \gtrsim q_{\rm Nep}$, the observed planet distribution is smooth and decreasing with a power-law index of $n = -0.85$ \citepalias[as in $dN / d\log q \propto q^n$ for $q > q_{\rm br}$;][]{Suzuki16}. 
This distribution is inconsistent with population synthesis models using the runaway gas accretion process from core accretion theory \citep{IdaLin04, Mordasini09}, which predict few planets with mass ratios in the range $1 < q / 10^{-4} < 5$ \citep{Suzuki18}.
Microlensing gives strong evidence for a break in the mass-ratio distribution at $q_{\rm br} \sim 5.6 \times 10^{-5}$ \citepalias{Suzuki16, Jung18}. While the power-law index  below $q_{\rm br}$ (i.e., $p$ for $dN / d\log q \propto q^p$ for $q < q_{\rm br}$) is constrained to be positive, the value is not well measured. The best-fit value from \citetalias{Suzuki16} is $p = 0.47$ while \citetalias{Jung18} prefers a value of $p \sim 4.5$ with $p\sim 2.5$ -- $9$ allowed at 1$\sigma$.

{Measuring the value of $p$ (or showing the small mass-ratio distribution is inconsistent with a power-law) requires a survey with substantial sensitivity to planets with mass ratios $q < q_{\rm br}$, which is only possible from space.}

\subsection{How Does the {Planet} Distribution Vary with Semi-Major Axis?
\label{sec:q_semia}}

Comparing the microlensing mass-ratio distribution to that derived from \Kep\, by \citet{Pascucci18} suggests the form of this distribution depends on separation from the host star. \citet{Pascucci18} find that the mass-ratio distribution of \Kep\, planets
with $P < 100\,$ days and $1 < r_{\rm p} / R_{\oplus} < 6$ ($5 < q / 10^{-5} < 8$) is also consistent with a broken power law with $q_{\rm br} \sim 3 \times 10^{-5}$. Their value of $q_{\rm br}$ is similar to the \citetalias{Jung18} value from microlensing, but the power-law indices are significantly different. They find $n = -2.9$ for planets with $q > q_{\rm br}$, which is much steeper than the microlensing value of $n = -0.86$ \citepalias{Suzuki16}. Likewise, the \citet{Pascucci18} value of $p \sim 1.0$ for planets with $q < q_{\rm br}$ is at the lower limit of what is allowed for microlensing planets \citepalias[$p \gtrsim 2.5$ 1-$\sigma$ lower limit and $p \gtrsim 0.$ at 3-$\sigma$][]{Suzuki16, Jung18}. These differences indicate the shape of mass-ratio distribution of planets changes with orbital period, but the details of this variation are virtually unconstrained. 

Measuring the value of $p$ for microlensing planets with $q < q_{\rm br}$ to enable a direct comparison to $p$ for short-period planets requires space-based microlensing. Furthermore, a space-based microlensing survey allows the detection of large planets over a much wider range of separations and thus a measurement of the variation of $n$ with separation. Finally, space-based microlensing can measure absolute projected separations for its planets, which will allow a more direct study of the variation with planetary orbital properties.

\subsection{How does the Planet Distribution Vary with Host Star Mass?
\label{sec:q_mstar}}

\citet{Pascucci18} find the mass-ratio distributions for M-, K-, and G-dwarf hosts are consistent with being drawn from the same population, i.e., that the forms of the broken power laws are consistent. Likewise, \citet{Udalski18} present tentative evidence that the break in the microlensing planet mass-ratio distribution is a break in mass-ratio rather than mass.
The lack of variation with host type suggests that mass ratio is a more fundamental quantity for defining planet populations than planet mass or radius, and therefore, that some aspect of planet formation scales with mass ratio for M-, K-, G-dwarfs. 

Only a space-based microlensing survey will routinely measure the masses of the lens stars, both those with and those without planets. Thus, a space-based microlensing survey can measure the dependence of the planet mass-ratio distribution on host mass for wider separation planets, and test whether mass-ratio is a more fundamental quantity than planet mass.

\subsection{Are There Free-Floating Planets and What is Their Mass Function?}

{Recent statistical results from the OGLE survey by \citep{Mroz17FFPs} find no evidence for a significant population of free-floating Jupiter-mass planets, and place an upper limit of such planets of 0.25 per star at 95\% confidence, refuting the previous claim of \citet{Sumi11}.}
However, \citet{Mroz17FFPs} do have tentative evidence for a population of smaller free-floating planets. Furthermore, \citet{Mroz18} and \citet{Mroz19} have identified {individual free-floating planets candidates} including one that is plausibly an Earth or Neptune.

{A microlensing survey from space can detect free-floating planets as small as $1 M_{\rm Mars}$. Furthermore, it can place stringent limits on potential stellar companions to demonstrate whether the population is truly free-floating (e.g., ejected) or consistent with being on wide, but bound orbits \citep{Gould16}.}

\section{Required Features of a Microlensing Survey from Space
\label{sec:requirements}}

\subsection{Detecting Small Planets}

\begin{figure}
\includegraphics[width=0.4\textwidth]{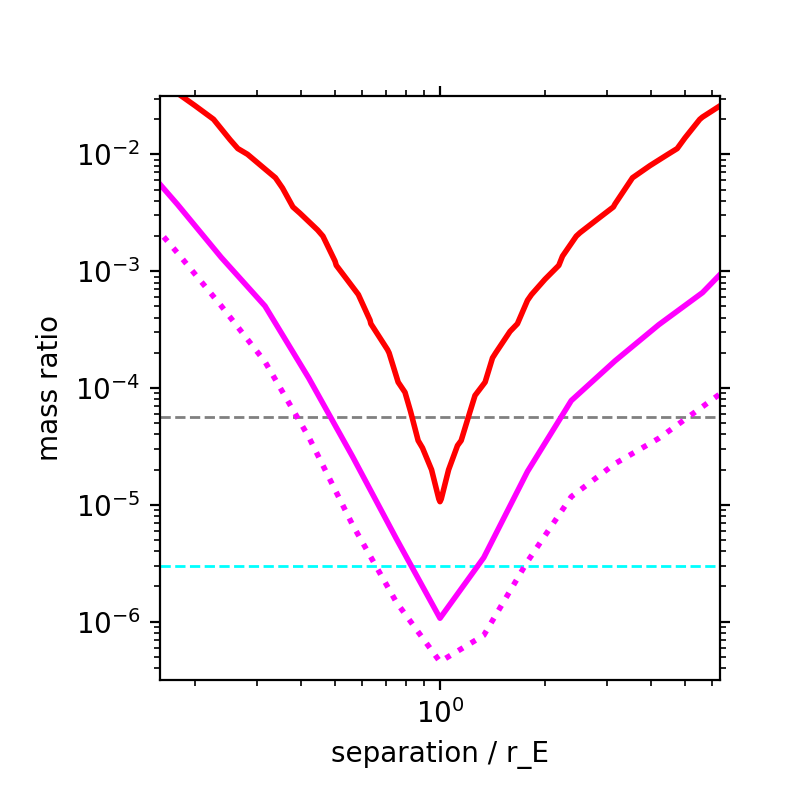}
\includegraphics[width=0.6\textwidth]{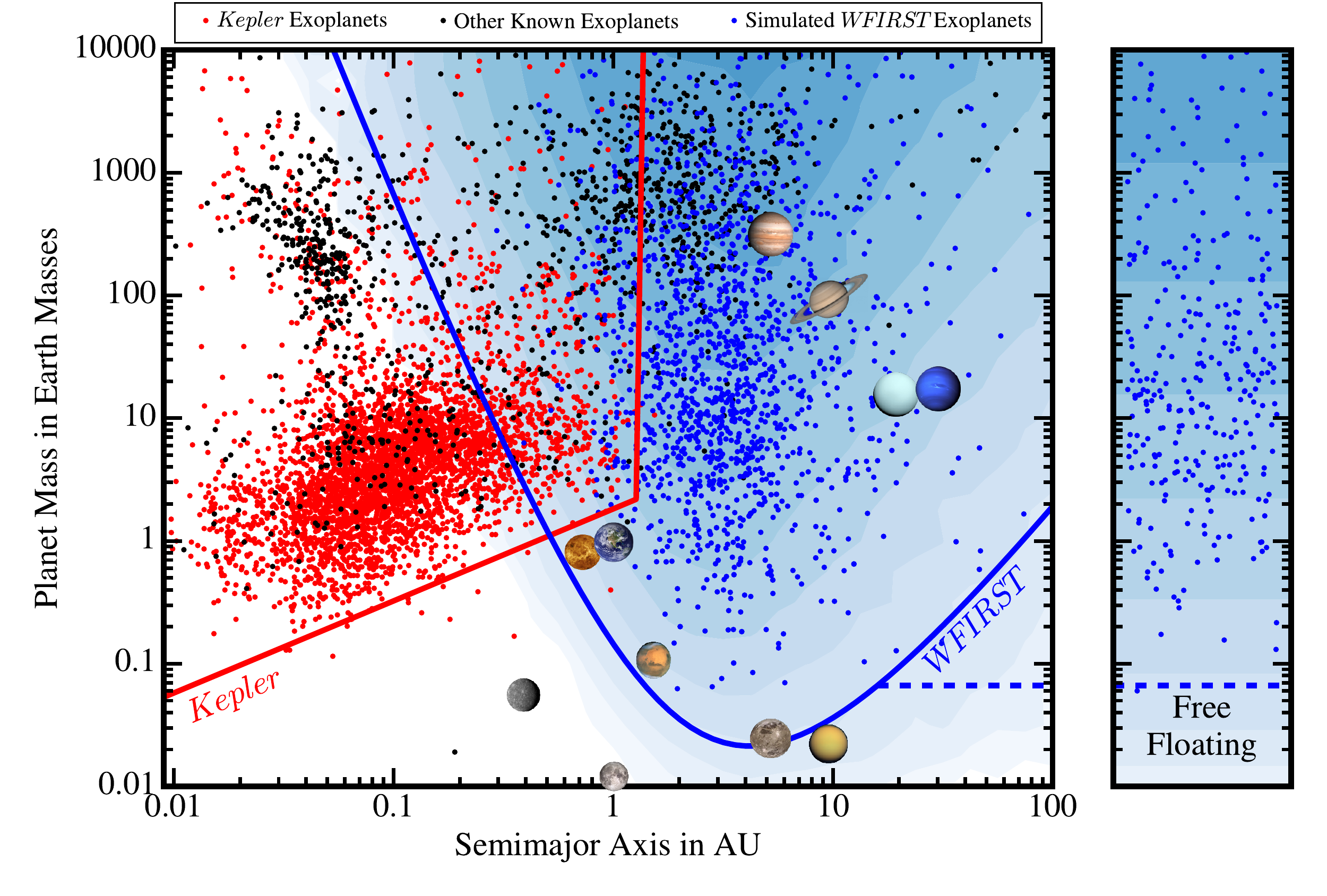}
\caption{
{\it Left}: \WF's precision photometry for $\sim 14,000$ events will measure the mass ratio function below $q_{\rm br}$ (gray dashed line). The solid and dashed magenta curves show where \WF\, can detect 20 and 5 planets per dex$^2$ if every star has such a planet. These curves extend substantially below $q_{\oplus}$ (cyan dashed line). In contrast, the red line shows for 20 planets per dex$^2$ for the ground-based MOA survey, which has limited leverage for measuring the mass-ratio distribution below $q_{\rm br}$.
{\it Right}: \WF\, is also able to measure masses for microlensing host stars and thus, their planets, as well as the orbital separations of those planets. The planet parameter space probed by \WF\, (blue) complements \Kep\, (red) as well as radial velocity, astrometry, and direct imaging.
\label{fig:sensitivities}}
\end{figure}

The smallest planets are challenging to detect because their
signals are intrinsically rare. While the dependence on planet mass is relatively shallow ($q^{1/2}$), the net effect is still that the geometric probability that an Earth mass-ratio planet produces a signal is $\lesssim 0.5\%$. Thus, detecting such small planets requires first detecting large numbers of microlensing events.

While ground-based microlensing surveys detect $\gtrsim 3,000$ events per year, these are biased toward the most numerous, and therefore, the faintest source stars. For ground-based, seeing-limited observations, many of these stars are below sky, and thus, the sky brightness sets the limit on the photometric precision that can be achieved. Furthermore, in seeing-limited observations, fainter stars are very likely to be blended with neighboring stars of comparable brightness, which dilutes the microlensing signal.

Diffraction-limited observations from space eliminate this problem giving high-precision photometry of even M-dwarf sources in the bulge. Such observations also limit blending with other stars.
Figure \ref{fig:sensitivities} shows the sensitivity of the MOA ground-based microlensing survey \citepalias{Suzuki16}, which is extremely limited below $q_{\rm br}$ (KMTNet is expected to do somewhat better than MOA because it has three sites and goes deeper, but its true performance is unknown because core survey operations are still underway).
By contrast, the substantial sensitivity of \WF\, \citep[adapted from simulations in][]{Penny18}. to planets with $q < q_{\rm br}$ will give a measurement of the power-law index of the mass-ratio function down to mass ratios smaller than $q_{\oplus}$. Even if no Earth mass-ratio planets are detected, this will give a strong constraint on the absence of such planets, especially compared to the measured frequency of Neptunes. 

Furthermore, the same properties that allow \WF\, to detect $q_{\oplus}$ planets allow it to detect planets near $q_{\rm br}$ at separations a factor of $\sim 2$ larger or smaller, i.e., a range from $\sim 0.3$--$100$ au (see Figure \ref{fig:sensitivities}). Thus, it will also be possible to study the behavior of {the mass-ratio function above $q_{\rm br}$} and the location of $q_{\rm br}$ as a function of separation.

\subsection{Host Masses and Distances $\rightarrow$ Planet Masses and Orbital Separations
\label{sec:req_masses}}

Answering the questions in Sections \ref{sec:q_semia} and \ref{sec:q_mstar} requires transforming from the microlensing observables (mass-ratio and separation as a fraction of the Einstein ring) to physical properties (planet mass and projected orbital separation). However,
it is difficult to measure the masses of the host stars for ground-based events. Although
it is often assumed that the typical microlensing host star is an M dwarf, in practice, the host stars with measured masses range from brown dwarfs \citep[e.g., ][]{Shvartzvald17_1195} to G-dwarfs \citep[e.g., ][]{Beaulieu16}. 
Thus, in the absence of host star mass measurements, the observed mass-ratio distribution is integrated across host mass, which will distort the distribution if it depends on host mass.  Measuring the host masses (and, therefore, also planet masses) for a large number of microlenses, 
would allow a measurement of both the direct planet mass distribution and the planet mass-ratio distribution for different types of host star. These results can then be unambiguously combined with results from other techniques to study dependence of the planet mass distribution on semi-major axis from $0$ to $>10$ au.

The challenge is that the mass of the host star $M_{\rm L}$ and its distance $D_{\rm L}$ are not fundamental microlensing observables. In fact, these properties are degenerate with each other and with a third property, the relative proper motion between the host (lens) star and the source star, $\mu_{\rm rel}$. Only one observable, the Einstein crossing time $t_{\rm E}$, is routinely measured in a microlensing event, meaning that two additional measurements are required to determine the host star mass and distance, and thus, the mass and physical separation of the planet (or non-planet).

In seeing-limited, ground-based observations, it is rare to measure the flux from the host star. Light
from additional stars is often blended into the PSF with the host and the source stars, and, even in the absence of blending, the host star light is usually too faint to be detected. For planets detected in ground-based observations, many of their host stars can be observed with high-resolution followup observations (either with adaptive optics or from space) to resolve blending and, if the observations are taken several years after the event, make a direct measurement of $\mu_{\rm rel}$ \citep[e.g., ][]{Bhattacharya18}. 
Thus, high-resolution observations alone can provide the two necessary additional measurements to determine the host star mass and distance from the Sun.

Even if $\mu_{\rm rel}$ is not measured,
i.e., light from the host and source stars is not separately resolved,
space-based observations of the combined source and host star measure or place upper limits on the host star flux (since the flux contribution from the source is known from the light curve model and the blending problem is resolved). Then, 
the host mass can still be measured if one additional microlensing effect is observed.
For planets, one of these effects is usually a measurement of the angular size of the Einstein ring $\theta_{\rm E} = \mu_{\rm rel} t_{\rm E}$, which can be derived from the finite source effect \citep[modulation of the observed magnification due to the finite size of the source star; ][]{Yoo04}. 
Another effect is the microlens parallax effect due to the motion of the observatory about the Sun. This physical effect is present in every microlensing light curve, but often at levels too small to be detected. 
However, the enhanced photometric precision from space-based observations means that it will be measurable for a much larger fraction of events.
Finally, diffraction-limited observations from space will be able to measure the astrometric microlensing effect (the shift in the centroid of the light due to the unequal images of the source) for some events, which will give an independent measurement of $\theta_{\rm E}$. 
Measuring this effect requires precision astrometric measurements while the event is ongoing, which is nearly impossible from the ground, except for black holes {\citep{GouldYee14, Lu16}}.

Thus, space-based microlensing offers many more opportunities to measure additional information that can be used to derive absolute masses for the host stars. While not all of these effects will be measurable in all events, 
mass measurements are expected for $> 40\%$ of planet hosts. Furthermore, some events will have measurements of several effects, making them over-constrained. This will allow internal tests of the various techniques that can be used to verify their uncertainties. 
Finally, these same measurements will yield distances to the host star.
These measurements permit the transformation from microlensing observables to physical parameters (left to right panels of Figure \ref{fig:sensitivities}) and measurements of variations in the planet population with semi-major axis and host star mass.

\subsection{Characterizing Free-Floating Planets
\label{sec:req_ffps}}

One of the major challenges for interpreting free-floating planets detected by microlensing is demonstrating that they are actually free-floating, i.e., that they do not have a host star. The microlensing light curve itself can be used to rule out stellar companions within a certain separation, i.e., place a lower limit on semi-major axis of the planet. However, for ground-based microlensing the limits tend to be modest, ruling out companions within a factor of two of $\sim 7$ au depending on the particular object. The increased photometric precision of a space-based microlensing survey would allow the detection of host stars at larger separations.
Furthermore, the high-resolution observations will place upper limits on the flux from a potential host star.

\vfill

\pagebreak

%\newpage
%\bibliography{/Users/jyee/Documents/tex/references}

\begin{thebibliography}{}
\expandafter\ifx\csname natexlab\endcsname\relax\def\natexlab#1{#1}\fi

\bibitem[{{Akeson} {et~al.}(2019){Akeson}, {Armus}, {Bachelet}, {Bailey},
  {Bartusek}, {Bellini}, {Benford}, {Bennett}, {Bhattacharya}, {Bohlin},
  {Boyer}, {Bozza}, {Bryden}, {Calchi Novati}, {Carpenter}, {Casertano},
  {Choi}, {Content}, {Dayal}, {Dressler}, {Dor{\'e}}, {Fall}, {Fan}, {Fang},
  {Filippenko}, {Finkelstein}, {Foley}, {Furlanetto}, {Kalirai}, {Gaudi},
  {Gilbert}, {Girard}, {Grady}, {Greene}, {Guhathakurta}, {Heinrich},
  {Hemmati}, {Hendel}, {Henderson}, {Henning}, {Hirata}, {Ho}, {Huff},
  {Hutter}, {Jansen}, {Jha}, {Johnson}, {Jones}, {Kasdin}, {Kelly}, {Kirshner},
  {Koekemoer}, {Kruk}, {Lewis}, {Macintosh}, {Madau}, {Malhotra}, {Mand el},
  {Massara}, {Masters}, {McEnery}, {McQuinn}, {Melchior}, {Melton},
  {Mennesson}, {Peeples}, {Penny}, {Perlmutter}, {Pisani}, {Plazas}, {Poleski},
  {Postman}, {Ranc}, {Rauscher}, {Rest}, {Roberge}, {Robertson}, {Rodney},
  {Rhoads}, {Rhodes}, {Ryan}, {Sahu}, {Sand}, {Scolnic}, {Seth}, {Shvartzvald},
  {Siellez}, {Smith}, {Spergel}, {Stassun}, {Street}, {Strolger}, {Szalay},
  {Trauger}, {Troxel}, {Turnbull}, {van der Marel}, {von der Linden}, {Wang},
  {Weinberg}, {Williams}, {Windhorst}, {Wollack}, {Wu}, {Yee}, \&
  {Zimmerman}}]{WFIRST2020s}
{Akeson}, R., {Armus}, L., {Bachelet}, E., {et~al.} 2019, arXiv e-prints,
  arXiv:1902.05569

\bibitem[{{Beaulieu} {et~al.}(2016){Beaulieu}, {Bennett}, {Batista}, {Fukui},
  {Marquette}, {Brillant}, {Cole}, {Rogers}, {Sumi}, {Abe}, {Bhattacharya},
  {Koshimoto}, {Suzuki}, {Tristram}, {Han}, {Gould}, {Pogge}, \&
  {Yee}}]{Beaulieu16}
{Beaulieu}, J.~P., {Bennett}, D.~P., {Batista}, V., {et~al.} 2016, \apj, 824,
  83

\bibitem[{{Bennett} \& {Rhie}(2002)}]{BennettRhie02}
{Bennett}, D.~P., \& {Rhie}, S.~H. 2002, \apj, 574, 985

\bibitem[{{Bennett} {et~al.}(2010){Bennett}, {Anderson}, {Beaulieu}, {Bond},
  {Cheng}, {Cook}, {Friedman}, {Gaudi}, {Gould}, {Jenkins}, {Kimble}, {Lin},
  {Mather}, {Rich}, {Sahu}, {Shao}, {Sumi}, {Tenerelli}, {Udalski}, \&
  {Yock}}]{Bennett10_MPF}
{Bennett}, D.~P., {Anderson}, J., {Beaulieu}, J.~P., {et~al.} 2010, arXiv
  e-prints, arXiv:1012.4486

\bibitem[{{Bhattacharya} {et~al.}(2018){Bhattacharya}, {Beaulieu}, {Bennett},
  {Anderson}, {Koshimoto}, {Lu}, {Batista}, {Blackman}, {Bond}, {Fukui},
  {Henderson}, {Hirao}, {Marquette}, {Mroz}, {Ranc}, \&
  {Udalski}}]{Bhattacharya18}
{Bhattacharya}, A., {Beaulieu}, J.~P., {Bennett}, D.~P., {et~al.} 2018, \aj,
  156, 289

\bibitem[{{Bonfils} {et~al.}(2013){Bonfils}, {Delfosse}, {Udry}, {Forveille},
  {Mayor}, {Perrier}, {Bouchy}, {Gillon}, {Lovis}, {Pepe}, {Queloz}, {Santos},
  {S{\'e}gransan}, \& {Bertaux}}]{Bonfils13}
{Bonfils}, X., {Delfosse}, X., {Udry}, S., {et~al.} 2013, \aap, 549, A109

\bibitem[{{Coughlin} {et~al.}(2016){Coughlin}, {Mullally}, {Thompson}, {Rowe},
  {Burke}, {Latham}, {Batalha}, {Ofir}, {Quarles}, {Henze}, {Wolfgang},
  {Caldwell}, {Bryson}, {Shporer}, {Catanzarite}, {Akeson}, {Barclay},
  {Borucki}, {Boyajian}, {Campbell}, {Christiansen}, {Girouard}, {Haas},
  {Howell}, {Huber}, {Jenkins}, {Li}, {Patil-Sabale}, {Quintana}, {Ramirez},
  {Seader}, {Smith}, {Tenenbaum}, {Twicken}, \& {Zamudio}}]{Coughlin16}
{Coughlin}, J.~L., {Mullally}, F., {Thompson}, S.~E., {et~al.} 2016, The
  Astrophysical Journal Supplement Series, 224, 12

\bibitem[{{Gould}(2016)}]{Gould16}
{Gould}, A. 2016, Journal of Korean Astronomical Society, 49, 123

\bibitem[{{Gould} \& {Yee}(2014)}]{GouldYee14}
{Gould}, A., \& {Yee}, J.~C. 2014, \apj, 784, 64

\bibitem[{{Ida} \& {Lin}(2004)}]{IdaLin04}
{Ida}, S., \& {Lin}, D.~N.~C. 2004, \apj, 604, 388

\bibitem[{{Jung} {et~al.}(2018){Jung}, {Gould}, {Zang}, {Hwang}, {Ryu}, {Han},
  {Yee}, {Albrow}, {Chung}, {Shin}, {Shvartzvald}, {Zhu}, {Cha}, {Kim}, {Kim},
  {Kim}, {Lee}, {Lee}, {Lee}, {Park}, {Pogge}, {Penny}, {Mao}, {Fouqu{\'e}}, \&
  {Wang}}]{Jung18}
{Jung}, Y.~K., {Gould}, A., {Zang}, W., {et~al.} 2018, arXiv e-prints,
  arXiv:1809.01288

\bibitem[{{Lagrange} {et~al.}(2009){Lagrange}, {Gratadour}, {Chauvin}, {Fusco},
  {Ehrenreich}, {Mouillet}, {Rousset}, {Rouan}, {Allard}, {Gendron}, {Charton},
  {Mugnier}, {Rabou}, {Montri}, \& {Lacombe}}]{Lagrange09}
{Lagrange}, A.-M., {Gratadour}, D., {Chauvin}, G., {et~al.} 2009, \aap, 493,
  L21

\bibitem[{{Lu} {et~al.}(2016){Lu}, {Sinukoff}, {Ofek}, {Udalski}, \&
  {Kozlowski}}]{Lu16}
{Lu}, J.~R., {Sinukoff}, E., {Ofek}, E.~O., {Udalski}, A., \& {Kozlowski}, S.
  2016, \apj, 830, 41

\bibitem[{{Mayor} {et~al.}(2011){Mayor}, {Marmier}, {Lovis}, {Udry},
  {S{\'e}gransan}, {Pepe}, {Benz}, {Bertaux}, {Bouchy}, {Dumusque}, {Lo Curto},
  {Mordasini}, {Queloz}, \& {Santos}}]{Mayor11}
{Mayor}, M., {Marmier}, M., {Lovis}, C., {et~al.} 2011, arXiv e-prints,
  arXiv:1109.2497

\bibitem[{{Mordasini} {et~al.}(2009){Mordasini}, {Alibert}, \&
  {Benz}}]{Mordasini09}
{Mordasini}, C., {Alibert}, Y., \& {Benz}, W. 2009, \aap, 501, 1139

\bibitem[{{Mr{\'o}z} {et~al.}(2017){Mr{\'o}z}, {Udalski}, {Skowron}, {Poleski},
  {Koz{\l}owski}, {Szyma{\'n}ski}, {Soszy{\'n}ski}, {Wyrzykowski},
  {Pietrukowicz}, {Ulaczyk}, {Skowron}, \& {Pawlak}}]{Mroz17FFPs}
{Mr{\'o}z}, P., {Udalski}, A., {Skowron}, J., {et~al.} 2017, \nat, 548, 183

\bibitem[{{Mr{\'o}z} {et~al.}(2018a){Mr{\'o}z}, {Ryu}, {Skowron}, {Udalski},
  {Gould}, {Szyma{\'n}ski}, {Soszy{\'n}ski}, {Poleski}, {Pietrukowicz},
  {Koz{\l}owski}, {Pawlak}, {Ulaczyk}, {OGLE Collaboration}, {Albrow}, {Chung},
  {Jung}, {Han}, {Hwang}, {Shin}, {Yee}, {Zhu}, {Cha}, {Kim}, {Kim}, {Kim},
  {Lee}, {Lee}, {Lee}, {Park}, {Pogge}, \& {KMTNet Collaboration}}]{Mroz18}
{Mr{\'o}z}, P., {Ryu}, Y.-H., {Skowron}, J., {et~al.} 2018, \aj, 155, 121

\bibitem[{{Mroz} {et~al.}(2018b){Mroz}, {Udalski}, {Bennett}, {Ryu}, {Sumi},
  {Shvartzvald}, {Skowron}, {Poleski}, {Pietrukowicz}, {Kozlowski},
  {Szymanski}, {Wyrzykowski}, {Soszynski}, {Ulaczyk}, {Rybicki}, {Iwanek},
  {Albrow}, {Chung}, {Gould}, {Han}, {Hwang}, {Jung}, {Shin}, {Yee}, {Zang},
  {Cha}, {Kim}, {Kim}, {Kim}, {Lee}, {Lee}, {Lee}, {Park}, {Pogge}, {Abe},
  {Barry}, {Bhattacharya}, {Bond}, {Donachie}, {Fukui}, {Hirao}, {Itow},
  {Kawasaki}, {Kondo}, {Koshimoto}, {Li}, {Matsubara}, {Muraki}, {Miyazaki},
  {Nagakane}, {Ranc}, {Rattenbury}, {Suematsu}, {Sullivan}, {Suzuki},
  {Tristram}, {Yonehara}, {Maoz}, {Kaspi}, \& {Friedmann}}]{Mroz19}
{Mroz}, P., {Udalski}, A., {Bennett}, D.~P., {et~al.} 2018, arXiv e-prints,
  arXiv:1811.00441

\bibitem[{{National Academies of Sciences Engineering and
  Medicine}(2018)}]{NAS_ESS18}
{National Academies of Sciences Engineering and Medicine}. 2018, Exoplanet
  Science Strategy (Washington, DC: The National Academies Press),
  doi:10.17226/25187

\bibitem[{{Pascucci} {et~al.}(2018){Pascucci}, {Mulders}, {Gould}, \&
  {Fernandes}}]{Pascucci18}
{Pascucci}, I., {Mulders}, G.~D., {Gould}, A., \& {Fernandes}, R. 2018, ArXiv
  e-prints, arXiv:1803.00777

\bibitem[{{Penny} {et~al.}(2018){Penny}, {Gaudi}, {Kerins}, {Rattenbury},
  {Mao}, {Robin}, \& {Calchi Novati}}]{Penny18}
{Penny}, M.~T., {Gaudi}, B.~S., {Kerins}, E., {et~al.} 2018, arXiv e-prints,
  arXiv:1808.02490

\bibitem[{{Perryman} {et~al.}(2014){Perryman}, {Hartman}, {Bakos}, \&
  {Lindegren}}]{Perryman14}
{Perryman}, M., {Hartman}, J., {Bakos}, G.~{\'A}., \& {Lindegren}, L. 2014,
  \apj, 797, 14

\bibitem[{{Shvartzvald} {et~al.}(2017){Shvartzvald}, {Yee}, {Calchi Novati},
  {Gould}, {Lee}, {Beichman}, {Bryden}, {Carey}, {Gaudi}, {Henderson}, {Zhu},
  {Spitzer Team}, {Albrow}, {Cha}, {Chung}, {Han}, {Hwang}, {Jung}, {Kim},
  {Kim}, {Kim}, {Lee}, {Park}, {Pogge}, {Ryu}, {Shin}, \& {KMTNet
  Group}}]{Shvartzvald17_1195}
{Shvartzvald}, Y., {Yee}, J.~C., {Calchi Novati}, S., {et~al.} 2017, \apj, 840,
  L3

\bibitem[{{Spergel} {et~al.}(2015){Spergel}, {Gehrels}, {Baltay}, {Bennett},
  {Breckinridge}, {Donahue}, {Dressler}, {Gaudi}, {Greene}, {Guyon}, {Hirata},
  {Kalirai}, {Kasdin}, {Macintosh}, {Moos}, {Perlmutter}, {Postman},
  {Rauscher}, {Rhodes}, {Wang}, {Weinberg}, {Benford}, {Hudson}, {Jeong},
  {Mellier}, {Traub}, {Yamada}, {Capak}, {Colbert}, {Masters}, {Penny},
  {Savransky}, {Stern}, {Zimmerman}, {Barry}, {Bartusek}, {Carpenter}, {Cheng},
  {Content}, {Dekens}, {Demers}, {Grady}, {Jackson}, {Kuan}, {Kruk}, {Melton},
  {Nemati}, {Parvin}, {Poberezhskiy}, {Peddie}, {Ruffa}, {Wallace}, {Whipple},
  {Wollack}, \& {Zhao}}]{Spergel15}
{Spergel}, D., {Gehrels}, N., {Baltay}, C., {et~al.} 2015, ArXiv e-prints,
  arXiv:1503.03757

\bibitem[{{Sumi} {et~al.}(2011){Sumi}, {Kamiya}, {Bennett}, {Bond}, {Abe},
  {Botzler}, {Fukui}, {Furusawa}, {Hearnshaw}, {Itow}, {Kilmartin}, {Korpela},
  {Lin}, {Ling}, {Masuda}, {Matsubara}, {Miyake}, {Motomura}, {Muraki},
  {Nagaya}, {Nakamura}, {Ohnishi}, {Okumura}, {Perrott}, {Rattenbury}, {Saito},
  {Sako}, {Sullivan}, {Sweatman}, {Tristram}, {Udalski}, {Szyma{\'n}ski},
  {Kubiak}, {Pietrzy{\'n}ski}, {Poleski}, {Soszy{\'n}ski}, {Wyrzykowski},
  {Ulaczyk}, \& {Microlensing Observations in Astrophysics (MOA)
  Collaboration}}]{Sumi11}
{Sumi}, T., {Kamiya}, K., {Bennett}, D.~P., {et~al.} 2011, \nat, 473, 349

\bibitem[{{Suzuki} {et~al.}(2016){Suzuki}, {Bennett}, {Sumi}, {Bond}, {Rogers},
  {Abe}, {Asakura}, {Bhattacharya}, {Donachie}, {Freeman}, {Fukui}, {Hirao},
  {Itow}, {Koshimoto}, {Li}, {Ling}, {Masuda}, {Matsubara}, {Muraki},
  {Nagakane}, {Onishi}, {Oyokawa}, {Rattenbury}, {Saito}, {Sharan}, {Shibai},
  {Sullivan}, {Tristram}, {Yonehara}, \& {MOA Collaboration}}]{Suzuki16}
{Suzuki}, D., {Bennett}, D.~P., {Sumi}, T., {et~al.} 2016, \apj, 833, 145

\bibitem[{{Suzuki} {et~al.}(2018){Suzuki}, {Bennett}, {Ida}, {Mordasini},
  {Bhattacharya}, {Bond}, {Donachie}, {Fukui}, {Hirao}, {Koshimoto},
  {Miyazaki}, {Nagakane}, {Ranc}, {Rattenbury}, {Sumi}, {Alibert}, \&
  {Lin}}]{Suzuki18}
{Suzuki}, D., {Bennett}, D.~P., {Ida}, S., {et~al.} 2018, \apj, 869, L34

\bibitem[{{Udalski} {et~al.}(2018){Udalski}, {Ryu}, {Sajadian}, {Gould},
  {Mr{\'o}z}, {Poleski}, {Szyma{\'n}ski}, {Skowron}, {Soszy{\'n}ski},
  {Koz{\l}owski}, {Pietrukowicz}, {Ulaczyk}, {Pawlak}, {Rybicki}, {Iwanek},
  {Albrow}, {Chung}, {Han}, {Hwang}, {Jung}, {Shin}, {Shvartzvald}, {Yee},
  {Zang}, {Zhu}, {Cha}, {Kim}, {Kim}, {Kim}, {Lee}, {Lee}, {Lee}, {Park},
  {Pogge}, {Bozza}, {Dominik}, {Helling}, {Hundertmark}, {J{\o}rgensen},
  {Longa-Pe{\~n}a}, {Lowry}, {Burgdorf}, {Campbell-White}, {Ciceri}, {Evans},
  {Figuera Jaimes}, {Fujii}, {Haikala}, {Henning}, {Hinse}, {Mancini},
  {Peixinho}, {Rahvar}, {Rabus}, {Skottfelt}, {Snodgrass}, {Southworth}, \&
  {von Essen}}]{Udalski18}
{Udalski}, A., {Ryu}, Y.-H., {Sajadian}, S., {et~al.} 2018, ArXiv e-prints,
  arXiv:1802.02582

\bibitem[{{Yoo} {et~al.}(2004){Yoo}, {DePoy}, {Gal-Yam}, {Gaudi}, {Gould},
  {Han}, {Lipkin}, {Maoz}, {Ofek}, {Park}, {Pogge}, {Szyma{\'n}ski}, {Udalski},
  {Szewczyk}, {Kubiak}, {{\.Z}ebru{\'n}}, {Pietrzy{\'n}ski}, {Soszy{\'n}ski},
  \& {Wyrzykowski}}]{Yoo04}
{Yoo}, J., {DePoy}, D.~L., {Gal-Yam}, A., {et~al.} 2004, \apj, 616, 1204

\bibitem[{Zhang {et~al.}(2018)Zhang, Zhu, Huang, Guzm{\'{a}}n, Andrews,
  Birnstiel, Dullemond, Carpenter, Isella, P{\'{e}}rez, Benisty, Wilner,
  Baruteau, Bai, \& Ricci}]{Zhang18}
Zhang, S., Zhu, Z., Huang, J., {et~al.} 2018, The Astrophysical Journal, 869,
  L47

\end{thebibliography}

\end{document}